\documentclass[prl,twocolumn]{revtex4-2}

\usepackage{graphicx}
\usepackage{dcolumn}
\usepackage{bm}

\begin{document}


\title{Measuring Density Functional Parameters from Electron Diffraction Patterns}

\author{Ding Peng}
\altaffiliation[Also at ]{Department of Physics, Norwegian University of Science and Technology (NTNU), Trondheim, Norway}
\email{Ding.Peng@monash.edu}
\author{Philip N. H. Nakashima}
\email{Philip.Nakashima@monash.edu}
\affiliation{Department of Materials Science and Engineering, Monash University, Victoria 3800, Australia}

\date{\today}

\begin{abstract}
We have integrated density functional theory (DFT) into quantitative convergent-beam electron diffraction (QCBED) to create a synergy between experiment and theory called \emph{QCBED-DFT}.  This synergy resides entirely in the electron density which, in real materials, gives rise to the experimental CBED patterns used by \emph{QCBED-DFT} to refine DFT model parameters.  We used it to measure the Hubbard energy, $U$, for two strongly correlated electron systems, NiO and CeB$_{6}$ ($U_{NiO} = 7.4 \pm 0.6$ eV and $U_{CeB_{6}} = 3.0 \pm 0.6$ eV), and the boron position parameter, $x$, for CeB$_{6}$ ($x = 0.1992 \pm 0.0003$).  In verifying our measurements, we demonstrate an accuracy test for any modelled electron density.
\end{abstract}

\maketitle

Density functional theory (DFT) is ubiquitous in materials science because it models electron densities and all resultant materials properties with useful accuracy at reasonable computational cost.  The theory is exact -- the exact functional of the exact density will give the exact energy of the system \cite{KohnNobel,JonesDFT,MedvedevDFT,KorthDFT} -- however, the exact functional is undefined.  Approximating it with functionals parametrized to reproduce energies of well-characterized systems has raised concerns that DFT \emph{“is straying from the path toward the exact functional”} \cite{MedvedevDFT,KorthDFT} because reproducing the correct energies does not imply reproduction of the true electron density \cite{KohnNobel,JonesDFT,MedvedevDFT,KorthDFT}.

Quantitative convergent-beam electron diffraction (QCBED) is an experimental pattern-matching technique, unsurpassed in accuracy and precision when measuring electron densities \cite{ZuoQCBED1988,BirdQCBED1992,SpenceQCBEDreview,ZuoetalQCBED1993,ZuoQCBED1993,DeiningerQCBED1994,SaundersQCBED1995,TsudaQCBED1995,HolmestadQCBED1995,ZuoWeickenmeierQCBED1995,PengQCBED1995,SaundersQCBED1996,MidgleyQCBED1996,BirkelandQCBED1996,ZuoQCBED1997,ZuoQCBEDNature,SaundersQCBEDI1999,SaundersQCBEDII1999,TsudaQCBED1999,DudarevDFTQCBED2000,StreltsovQCBED2001,TsudaQCBED2002,JiangQCBEDI2003,FriisQCBEDI2003,StreltsovQCBED2003,JiangQCBEDII2003,FriisQCBEDII2003,OgataQCBED2004,JiangQCBED2004,FriisQCBED2004,ZuoBonding2004,NakaQCBED2005,FriisQCBED2005,NakaTDQCBED2007,TsudaQCBED2010,SangQCBEDI2010,SangQCBEDII2010,NakaQCBED2010,NakaQCBEDScience,MidgleyPerspective,SangQCBED2011,SangQCBED2012,NakaNoise2012,SangQCBEDI2013,SangQCBEDII2013,Peng2017,NakaQCBEDReview2017,ZuoSpence2017,QCrReview}.  We have integrated DFT into QCBED to fit CBED patterns calculated from DFT-modelled electron densities to experimental CBED patterns from the actual electron density in materials.  Our method, \emph{QCBED-DFT}, refines DFT model parameters without comparing energies or properties at all, confining the refinements to electron densities alone.

The electron density, $\rho(\bf{r})$, is the dominant determinant of materials properties and, if accurately known by other means, can provide a 3-dimensional constraint for DFT, as emphasized by Kohn in his Nobel Lecture \cite{KohnNobel}.  In this spirit, X-ray diffraction experiments have recently been applied in frozen-density embedding theory (FDET) to constrain the Hohenberg-Kohn functional \cite{RicardiFDET2020} and analogs in many-body wave-function methods have given rise to the well-established field of X-ray-constrained wave-functions (XCW) \cite{JayatilakaPRL,JayatilakaXCWI2001,JayatilakaXCWII2001,ChecinskaXCW2013,GenoniXCW2017,WoinskaXCW2017,QCrReview}.

For both XCW and FDET, the accuracy of the X-ray diffraction data is critical.  Extinction \cite{CoppensXCD1997} can be problematic and is the result of multiple scattering in the context of a kinematic scattering analysis.  This is not an issue for QCBED because the strongly dynamical nature of electron diffraction necessitates a full dynamical scattering treatment.  It also gives rise to the very detailed intensity distributions in CBED patterns (see Figs. \ref{CQCBEDfig} and \ref{QCBED-DFTfig}; Tables S4 and S8 in the Supplemental Materials \cite{SupplMats}) that QCBED typically matches at the rate of $\sim10^{3}$ independent intensities per reflection.  This leads to parameter oversampling by approximately 3 orders of magnitude.  Furthermore, the routine ability to position nanometer-sized electron probes with sub-nanometer spatial precision in a transmission electron microscope (TEM) means that very small volumes of perfect crystal ($\sim10^{9}$ times smaller than in X-ray diffraction) can be selected for acquiring CBED patterns.  All of these factors contribute to the very high precision and accuracy in electron density measurements for which QCBED is renowned \cite{ZuoQCBED1988,BirdQCBED1992,SpenceQCBEDreview,ZuoetalQCBED1993,ZuoQCBED1993,DeiningerQCBED1994,SaundersQCBED1995,TsudaQCBED1995,HolmestadQCBED1995,ZuoWeickenmeierQCBED1995,PengQCBED1995,SaundersQCBED1996,MidgleyQCBED1996,BirkelandQCBED1996,ZuoQCBED1997,ZuoQCBEDNature,SaundersQCBEDI1999,SaundersQCBEDII1999,TsudaQCBED1999,DudarevDFTQCBED2000,StreltsovQCBED2001,TsudaQCBED2002,JiangQCBEDI2003,FriisQCBEDI2003,StreltsovQCBED2003,JiangQCBEDII2003,FriisQCBEDII2003,OgataQCBED2004,JiangQCBED2004,FriisQCBED2004,ZuoBonding2004,NakaQCBED2005,FriisQCBED2005,NakaTDQCBED2007,TsudaQCBED2010,SangQCBEDI2010,SangQCBEDII2010,NakaQCBED2010,NakaQCBEDScience,MidgleyPerspective,SangQCBED2011,SangQCBED2012,NakaNoise2012,SangQCBEDI2013,SangQCBEDII2013,Peng2017,NakaQCBEDReview2017,ZuoSpence2017,QCrReview}.

In QCBED, an experimental, point spread function (PSF)-corrected \cite{NakaJohnPSF2003} CBED pattern is fitted with a theoretically calculated one by refining the parameters to which the diffracted intensities are most sensitive.  Recent methods incorporate angular differentiation to remove most of the inelastic signal that impedes matching with an elastic electron scattering theory \cite{NakaQCBED2010,NakaQCBEDScience,MidgleyPerspective,NakaQCBEDReview2017,QCrReview}.  In Figs. \ref{CQCBEDfig} and \ref{QCBED-DFTfig}, where the differences between conventional QCBED (CQCBED) and \emph{QCBED-DFT} are illustrated, the experimental data are labelled “$I'_{exp’t}$” and the refined calculated intensities are marked “$I'_{calc.}$”.

The crystal potential, $V(\bf{r})$, is essential to the calculation of CBED patterns and its Fourier coefficients (structure factors), $V_{hkl}$, are exactly interconvertible with those of $\rho(\bf{r})$, i.e. $F_{hkl}$, by the Mott-Bethe formula \cite{Mott1930,Bethe1930}. In CQCBED, an independent atom model (IAM), which ignores interatomic bonding, is used and the refined parameters include the bonding-affected $V_{hkl}$.

The differences between the CQCBED-measured structure factors and their IAM counterparts constitute a set of difference structure factors which quantify the bonding potential, ${\Delta}V(\bf{r})$ (Fig. \ref{CQCBEDfig}).  This is accurate only if:  (i) all bonding-affected structure factors have been refined; (ii) the IAM represents isolated atoms accurately.

\begin{figure*}
\includegraphics[width=0.9\linewidth]{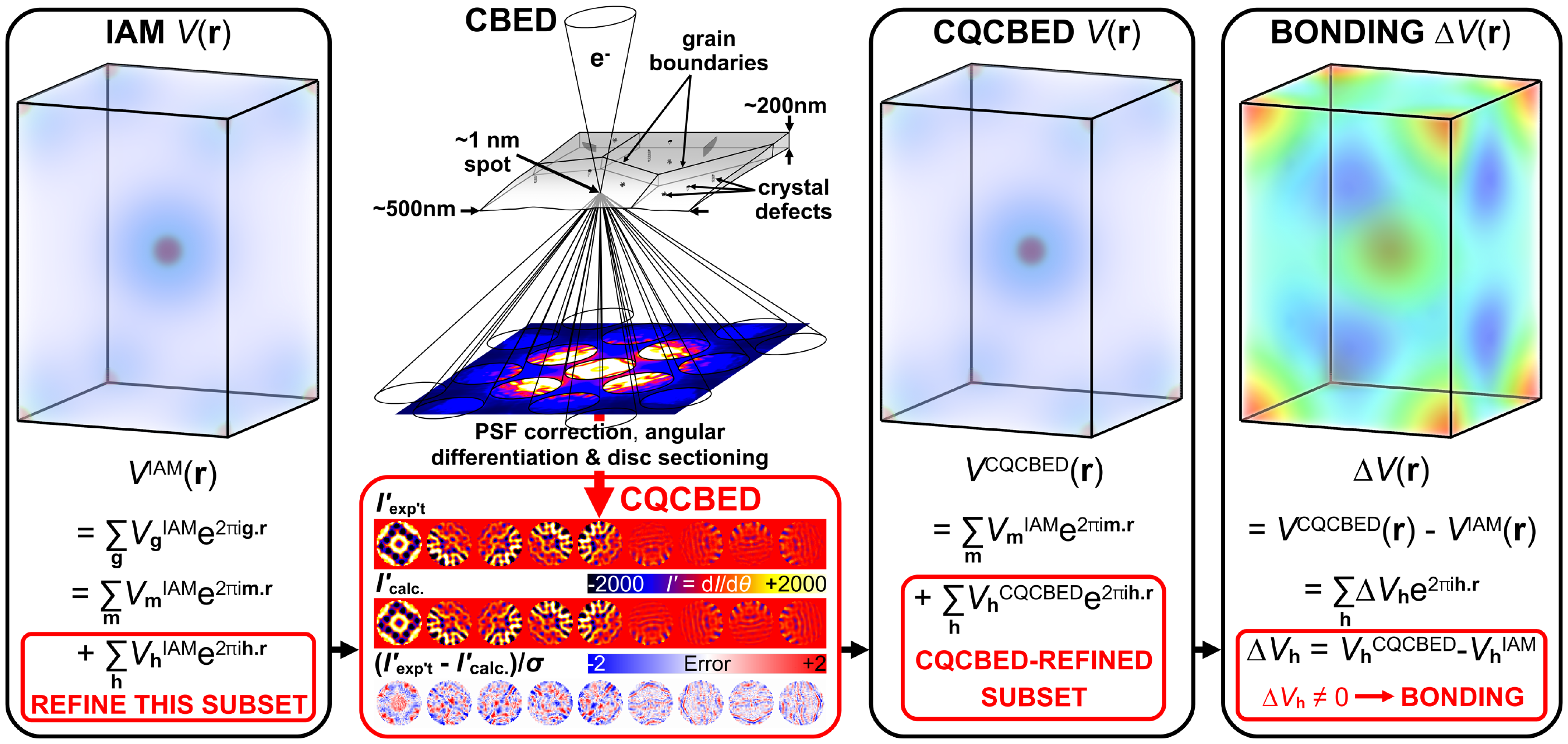}
\caption{(color online) Conventional QCBED (CQCBED).  A few parameters are adjusted to fit a calculated CBED pattern to an experimental one, including the structure factors, $V_{\bf{h}}$, to which the intensities are most sensitive.  These form a very small subset of all structure factors, $V_{\bf{g}}$, required for a full dynamical electron scattering calculation.  The unrefined remainder, $V_{\bf{m}}$, are obtained from an IAM, i.e. $V_{\bf{m}}^{IAM}$.  As the refinement progresses, the modified structure factors which start with IAM values, $V_{\bf{h}}^{IAM}$, take on new values, $V_{\bf{h}}^{CQCBED}$.  The bonding potential is the difference between the CQCBED-refined and IAM potentials, and is equivalent to computing the Fourier sum using the difference structure factors, ${\Delta}V_{\bf{h}}$.  Here, the bonding potential in aluminum from \cite{NakaQCBEDScience} has been used as an example of CQCBED.}
\label{CQCBEDfig}
\end{figure*}

\begin{figure*}
\includegraphics[width=0.9\linewidth]{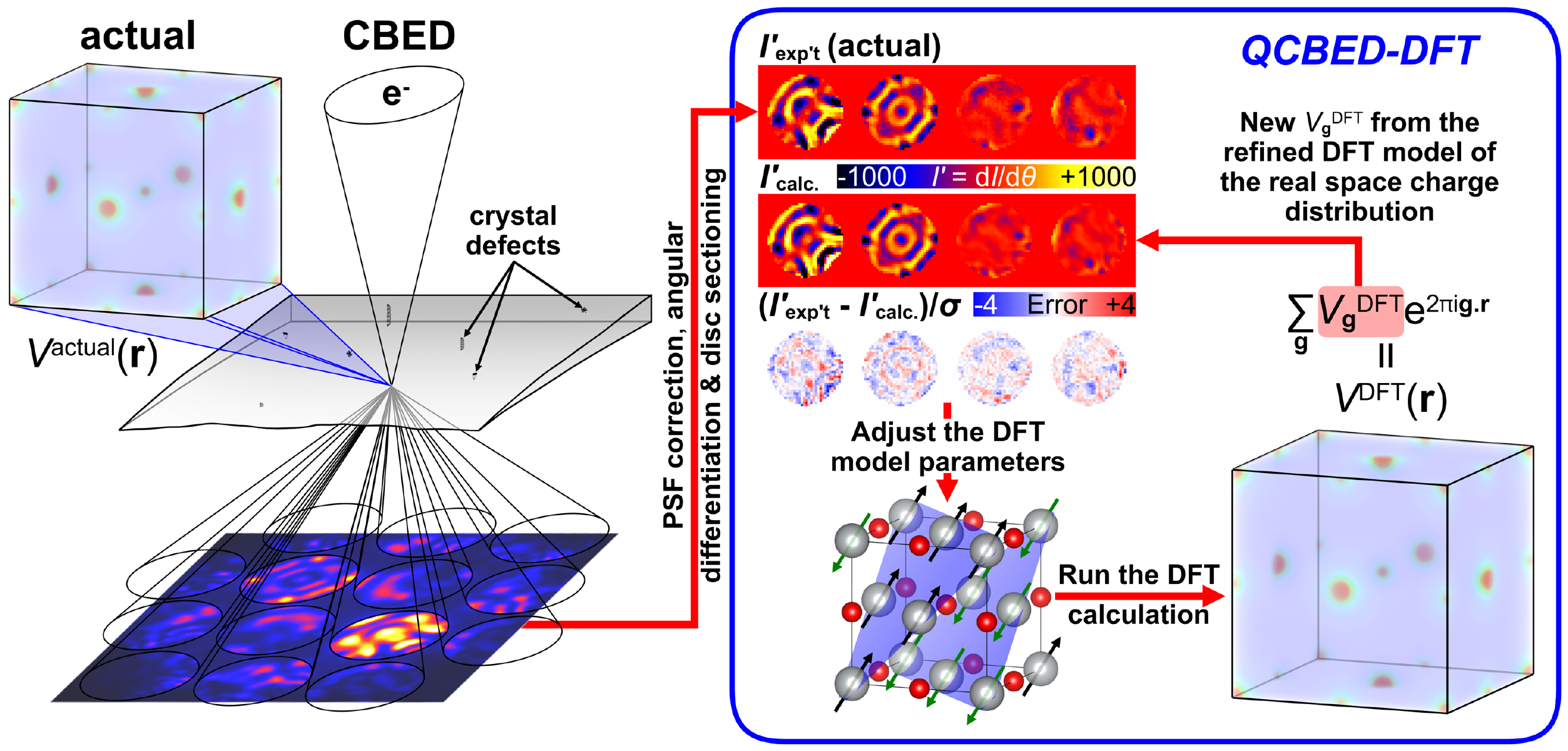}
\caption{(color online) \emph{QCBED-DFT}.  The DFT model parameters are refined which, when adjusted, change the simulated electron density in real space, thereby changing \emph{all} structure factors, $V_{\bf{g}}^{DFT}$, of the crystal potential returned to the electron scattering calculations.  By optimizing the fit between the calculated and experimental CBED patterns, the DFT model parameters are refined (see \cite{SupplMats} for details).  The present example involves antiferromagnetic NiO -- a subject of this work.}
\label{QCBED-DFTfig}
\end{figure*}

\begin{figure*}
\includegraphics[width=0.9\linewidth]{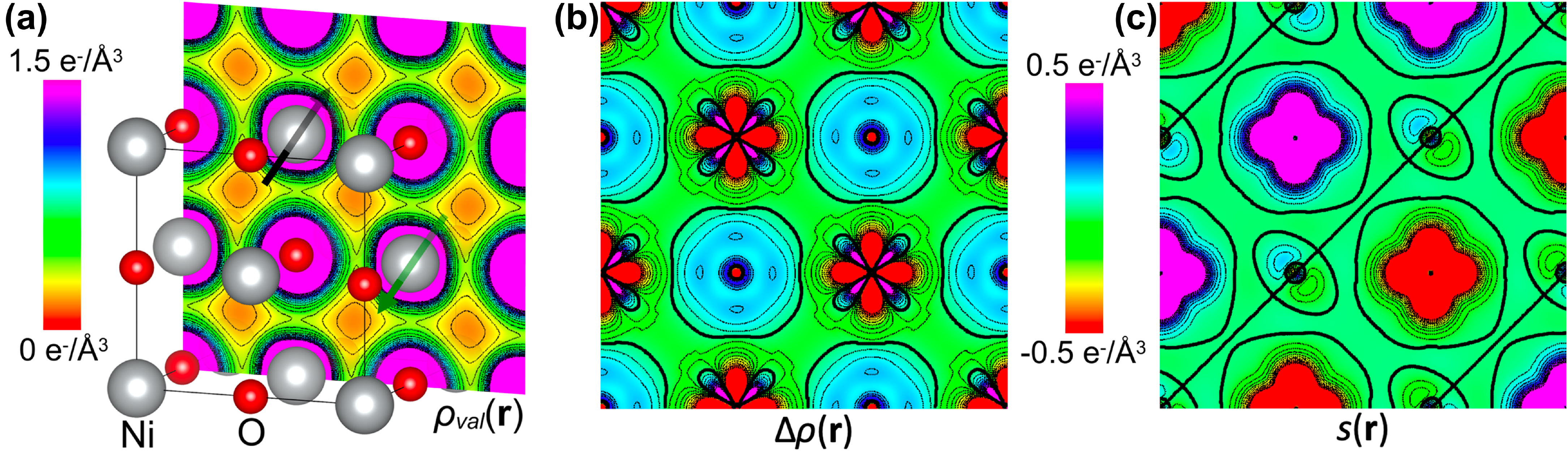}
\caption{(color online) The electronic structure of NiO determined by \emph{QCBED-DFT} with $U$ = 7.4 eV and $J$ = 0.95 eV \cite{DudarevDFTQCBED2000,AnisimovHubbardU,DudarevNiO} at $T$ = 0 K.  The valence electron density, ${\rho}_{val}(\bf{r})$, is mapped in the (020) plane with a contour interval of 0.1 e$^{-}$Å$^{-3}$ (a).  Nearest-neighbor nickel atoms (grey) have opposite spins indicated by the black and green arrows parallel and antiparallel to [112].  The bonding and electron spin densities, ${\Delta}{\rho}(\bf{r})$ (b) and $s(\bf{r})$ (c) respectively, are plotted for the same region as ${\rho}_{val}(\bf{r})$ (a), but with a contour interval of 0.05 e$^{-}$Å$^{-3}$ and the zero contour thickened.}
\label{NiOfig}
\end{figure*}

By replacing the IAM with DFT, \emph{QCBED-DFT} treats the material as an ensemble of bonded atoms, not independent ones.  Instead of refining small subsets of structure factors, \emph{QCBED-DFT} refines DFT model parameters, altering the simulated electron density in real space and therefore changing \emph{all} structure factors used to calculate CBED patterns (Fig. \ref{QCBED-DFTfig}).  At present, \emph{QCBED-DFT} uses Bloch wave code \cite{ZuoQCBED1993,BetheBlochWaves} for the CBED intensity calculations and calls \emph{Wien2k} \cite{BlahaWien2k} for the DFT-calculated crystal potential as input (see \cite{SupplMats} for details).

Instead of testing density functionals by comparing system energies and materials properties, \emph{QCBED-DFT} interrogates ${\rho}(\bf{r})$ directly because the experimental CBED patterns being matched are a direct consequence of $V(\bf{r})$, and thus ${\rho}(\bf{r})$, in the actual material.

While QCBED experimentally constrains DFT parameters, the $V(\bf{r})$ calculated by DFT at each iteration of \emph{QCBED-DFT} represents the material better than an IAM, increasing the accuracy of QCBED.  Furthermore, DFT provides the theoretical framework for translating the \emph{QCBED-DFT}-optimized electron density into a large suite of materials properties, assuming that the functional and calculation protocols are sufficiently accurate for the material being modelled \cite{JonesDFT,QCrReview,AdlerCorrelMats,LejaeghereScience2016,TranDFT,BlahaWien2k}.

Strongly correlated electron systems have challenged DFT \cite{JonesDFT,AnisimovHubbardU,AdlerCorrelMats} and one way of dealing with electron correlations has been to add the Hubbard energy parameter, $U$, into the exchange-correlation functional \cite{JonesDFT,AnisimovHubbardU,AdlerCorrelMats}.  In our investigations of NiO and CeB$_{6}$, we used the Perdew-Burke-Ernzerhof (PBE) generalized gradient approximation (GGA) \cite{BlahaWien2k,PerdewGGA} in a GGA(PBE) + $U$ configuration and we refined $U$.  This was done within the augmented plane wave plus local orbitals (APW + lo) regime of \emph{Wien2k} \cite{BlahaWien2k}.  For CeB$_{6}$, we also refined the positions of the boron atoms because of significant discrepancies in the literature \cite{SatoCeB6,EliseevCeB6,BlombergCeB6,StreltsovCeB6,TanakaCeB6}.  All experimental details, DFT and QCBED settings within \emph{QCBED-DFT}, refinement outputs, materials properties and electronic structure morphologies are discussed in \cite{SupplMats}.

Figure \ref{QCBED-DFTfig} shows a typical example of a \emph{QCBED-DFT} refinement from the nine that were performed for NiO (see Tables S1-S4 in \cite{SupplMats}).  All of them involved the refinement of only 8 parameters, including the Hubbard energy, $U$, to fit 3,026 independent intensities per pattern.  The nine CBED patterns came from regions of different specimen thickness ranging from 1,288 Å to 2,001 Å (see Tables S1-S3 in \cite{SupplMats}) and were collected with 202.7 ± 0.2 keV electrons incident near $<$011$>$.  We report a value of $U$ = 7.4 ± 0.6 eV for NiO from our \emph{QCBED-DFT} refinements, with a Hund exchange parameter of $J$ = 0.95 eV \cite{DudarevDFTQCBED2000,AnisimovHubbardU,DudarevNiO}.  This result is within the range of $U$ = 4.6 eV - 8.0 eV reported previously \cite{AnisimovHubbardU,DudarevNiO,AnisimovNiO,WeiNiO,HugelNiO,KwonNiO,MadsenNiO,CococcioniNiO,CaiNiO}.

Figure \ref{NiOfig} presents the \emph{QCBED-DFT}-optimized electronic structure of NiO at $T$ = 0 K.  Figures \ref{NiOfig}a - \ref{NiOfig}c show that the valence and bonding electron densities, ${\rho}_{val}(\bf{r})$ (a) and ${\Delta}{\rho}(\bf{r})$ (b) respectively, are insensitive to the opposed magnetic moments of nearest-neighbor nickel atoms whilst the electron spin density, $s(\bf{r})$ (c), distinguishes them very clearly, as expected.

Whilst CQCBED can determine ${\Delta}{\rho}(\bf{r})$ accurately (given an accurate IAM), the determination of ${\rho}_{val}(\bf{r})$ would require an accurate independent ion model.  On its own, CQCBED cannot determine $s(\bf{r})$ because CBED is insensitive to electron spin.  Revealing ${\rho}_{val}(\bf{r})$ and $s(\bf{r})$ is a benefit of the theoretical framework provided by DFT in the context of the electron density born out of that same framework.  However, in contrast to stand-alone DFT, the electron density in the present case has been refined against experimental CBED patterns which are a direct consequence of the actual ${\rho}(\bf{r})$ in the real material.

\begin{figure*}
\includegraphics[width=0.9\linewidth]{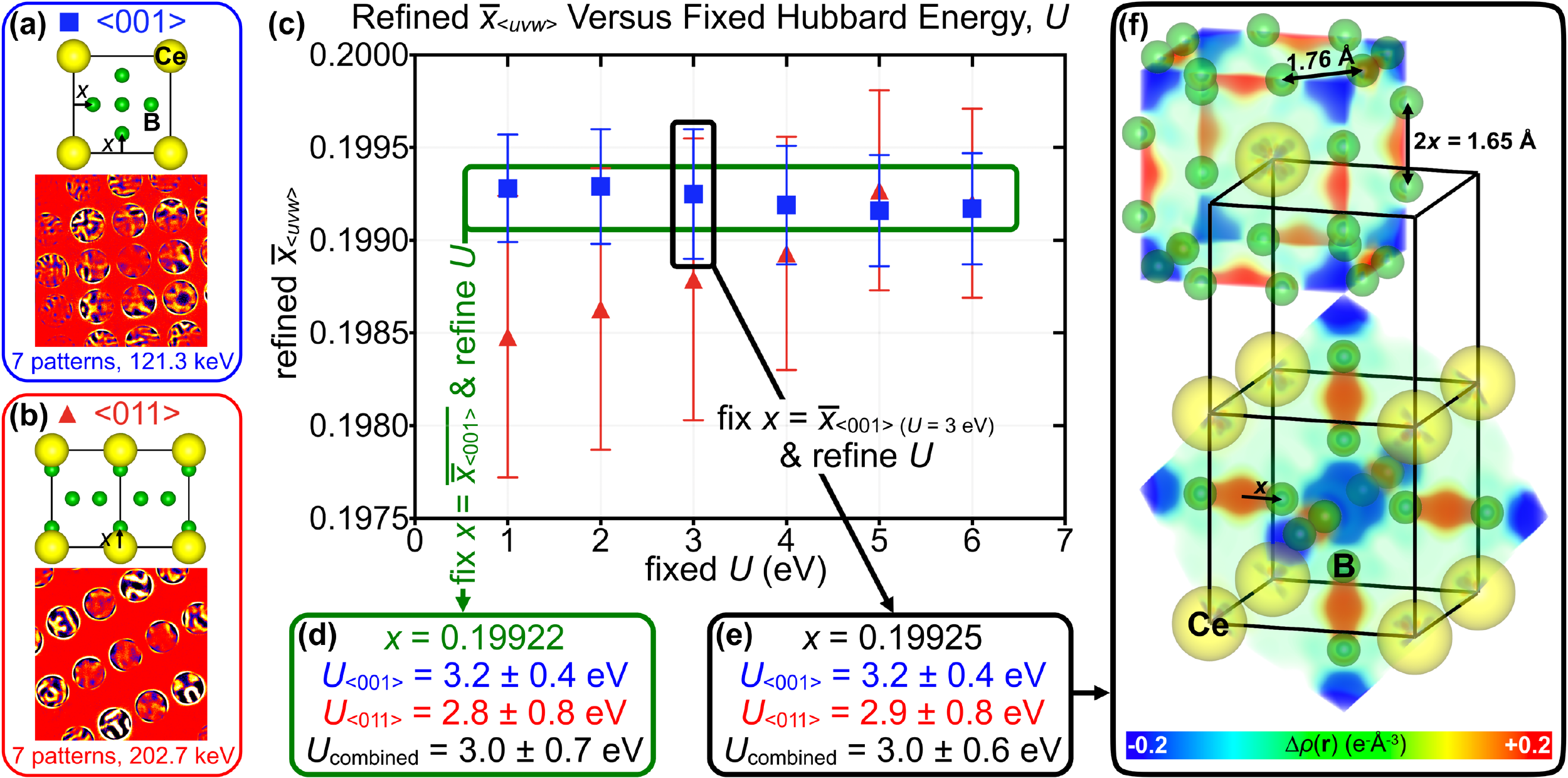}
\caption{(color online) Determination of the atomic structure parameter, $x$, the Hubbard energy, $U$, and ${\Delta}{\rho}(\bf{r})$ for CeB$_{6}$ using \emph{QCBED-DFT}.  The boron atom positions are defined by $x$ (a, b, f).  Sets of seven CBED patterns from near $<$001$>$ (a) and $<$011$>$ (b) were matched by fixing $U$ and refining $x$.  The mean values of $x$ from each zone axis, $\bar{x}_{<uvw>}$, are graphed with their uncertainties (c) for each value of $U$ (see Tables S6 and S7 in \cite{SupplMats}).  Fixing $x$ = $\overline{\bar{x}_{<001>}}$ (the mean of $\bar{x}_{<001>}$ over all fixed values of $U$), $U$ was then refined and results for the $<$001$>$ CBED data ($U_{<001>}$), the $<$011$>$ data ($U_{<011>}$) and all data combined ($U_{combined}$) are reported (d).  This was repeated for $x$ = $\bar{x}_{<001>}$ ($U$ = 3 eV) = 0.19925 (e).  The final results (e) give rise to the plot of ${\Delta}{\rho}(\bf{r})$ at $T$ = 0 K (f).  The top half plots ${\Delta}{\rho}(\bf{r})$ within a unit cell centered on a cerium atom and the bottom half ranges over 2 x 2 x 2 unit cells centered on a boron octahedron and truncated by \{011\} planes.}
\label{CeB6fig}
\end{figure*}

For CeB$_{6}$, in addition to refining $U$, we also refined the boron position parameter, $x$ (Figs. \ref{CeB6fig}a, b and f) because of significant discrepancies in previously reported values, i.e. $x$ = 0.200 ± 0.002 \cite{SatoCeB6,EliseevCeB6,BlombergCeB6,StreltsovCeB6,TanakaCeB6} (see Table S5 in \cite{SupplMats}).

We analysed 14 CBED patterns with \emph{QCBED-DFT}, 7 collected with 121.3 ± 0.2 keV electrons incident near $<$001$>$ and 7 collected with 202.7 ± 0.2 keV electrons incident near $<$011$>$.  All patterns came from regions of different specimen thickness ranging from 1,186 Å to 1,719 Å (see Tables S6 and S7 in \cite{SupplMats}).  This experimental variety tests the precision of the results more robustly than repeated measurements under the same conditions.

Our refinements of $x$ and $U$ involved two stages with only one parameter refined during each stage in order to assess and mitigate parameter correlations.

In stage one, all 14 CBED patterns were matched by refining only $x$, while $U$ was fixed at values from 1 eV to 6 eV in increments of 1 eV.  The mean $x$ obtained from the 7 refinements near each zone axis, $\bar{x}_{<uvw>}$, is plotted for each fixed value of $U$, for each of $<$001$>$ and $<$011$>$, in Fig. \ref{CeB6fig}c.  The independence of $x$ from $U$ is evident in all refinements using CBED patterns near $<$001$>$ whilst they appear to be linearly correlated in the refinements using near $<$011$>$ data.  This is probably due to the manifestation of $x$ in two non-colinear directions perpendicular to $<$001$>$ (Fig. \ref{CeB6fig}a) as opposed to just one such appearance perpendicular to $<$011$>$ (Fig. \ref{CeB6fig}b).  Furthermore, all boron atom columns in $<$001$>$ projections are well separated from cerium columns, whilst this is not the case for $<$011$>$.  As a first estimate of $x$, we used only the results from the $<$001$>$ data, averaging over $\bar{x}_{<001>}$ for all fixed values of $U$ to get $\overline{\bar{x}_{<001>}}$ = 0.19922.

Stage two involved fixing $x$ = 0.19922 and refining $U$ using all 14 CBED patterns.  The results are summarized in Fig. \ref{CeB6fig}d and suggest that $U$ = 3 eV.  We then repeated the refinements of $U$ with $x$ fixed at $\bar{x}_{<001>}$ ($U$ = 3 eV), i.e. $x$ = 0.19925.  The results of this final set of refinements are summarized in Fig. \ref{CeB6fig}e and are not significantly different to the preceding results in Fig. \ref{CeB6fig}d.

We conclude that $x$ = $\bar{x}_{<001>}$ ($U$ = 3 eV) = 0.1992 ± 0.0003 and $U$ = 3.0 ± 0.6 eV.  This is in agreement with Sato ($x$ = 0.19923 ± 0.00006) \cite{SatoCeB6}, Blomberg \emph{et al.} ($x$ = 0.1992 ± 0.0001) \cite{BlombergCeB6} and Streltsov \emph{et al.} ($x$ = 0.1995 ± 0.0003) \cite{StreltsovCeB6} for $x$, whilst our refined value of $U$ agrees with the study of Barman \emph{et al.} \cite{BarmanCeB6}.

Applying $x$ = 0.19925 (Fig. \ref{CeB6fig}e, i.e. $\bar{x}_{<001>}$ ($U$ = 3 eV) without rounding) and $U$ = 3 eV, the resulting ${\Delta}{\rho}(\bf{r})$ at $T$ = 0 K is shown in Fig. \ref{CeB6fig}f.  A more detailed discussion is presented in \cite{SupplMats}; however, the key finding is that the boron atoms do not form octahedra by bonding in an octahedral configuration.  Instead, they bond in dumbbell pairs coordinated to nearest-neighbor cerium atoms and this is what results in the formation of boron octahedra.  It is notable that the B – B dumbbells have bond lengths of 1.65 Å, which are significantly shorter than the edge lengths of the boron octahedra (1.76 Å).  Furthermore, there is very strong anti-bonding electron density (${\Delta}{\rho}(\bf{r})$ $<$ 0) within each boron octahedron.

A priority of quantum crystallography \cite{QCrReview} is to arrive at more accurate electron densities, which raises an important question:  \emph{How can one test the accuracy of a modelled ${\rho}(\bf{r})$?}.  The answer, at least in the context of this work, is in the form of CQCBED (Fig. \ref{CQCBEDfig}).

In CQCBED, discrepancies between the modelled ${\rho}(\bf{r})$ and the actual ${\rho}(\bf{r})$ from which the experimental CBED patterns arise, will cause individual structure factors to change from the modelled values in order to minimize the pattern mismatch.  This is how CQCBED with an IAM has been applied to date -- to measure changes in structure factors from IAM values due to bonding (Fig. \ref{CQCBEDfig}) \cite{ZuoQCBED1988,BirdQCBED1992,SpenceQCBEDreview,ZuoetalQCBED1993,ZuoQCBED1993,DeiningerQCBED1994,SaundersQCBED1995,TsudaQCBED1995,HolmestadQCBED1995,ZuoWeickenmeierQCBED1995,PengQCBED1995,SaundersQCBED1996,MidgleyQCBED1996,BirkelandQCBED1996,ZuoQCBED1997,ZuoQCBEDNature,SaundersQCBEDI1999,SaundersQCBEDII1999,TsudaQCBED1999,DudarevDFTQCBED2000,StreltsovQCBED2001,TsudaQCBED2002,JiangQCBEDI2003,FriisQCBEDI2003,StreltsovQCBED2003,JiangQCBEDII2003,FriisQCBEDII2003,OgataQCBED2004,JiangQCBED2004,FriisQCBED2004,ZuoBonding2004,NakaQCBED2005,FriisQCBED2005,NakaTDQCBED2007,TsudaQCBED2010,SangQCBEDI2010,SangQCBEDII2010,NakaQCBED2010,NakaQCBEDScience,MidgleyPerspective,SangQCBED2011,SangQCBED2012,NakaNoise2012,SangQCBEDI2013,SangQCBEDII2013,Peng2017,NakaQCBEDReview2017,ZuoSpence2017,QCrReview}.  If the IAM in CQCBED were replaced with the true ${\rho}(\bf{r})$, then the refinement of individual structure factors would result in no changes from their modelled values.  We used this principle to test the accuracy of the \emph{QCBED-DFT}-determined ${\rho}(\bf{r})$.

We performed two sets of CQCBED refinements for NiO and CeB$_{6}$.  In the first set, we used the $V(\bf{r})$ (and thus ${\rho}(\bf{r})$) resulting from our \emph{QCBED-DFT} refinements and we label this CQCBED-DFT for the purposes of the present discussion.  For comparison, we also performed a second set of CQCBED refinements using the standard IAM \cite{DoyleTurnerIAM,IntTabCryst1974,IntTabCryst2006} for $V(\bf{r})$ and thus ${\rho}(\bf{r})$.  We label this set of refinements CQCBED-IAM.  The individually refined structure factors in CQCBED-DFT and CQCBED-IAM were not only those to which the experimental CBED patterns were most sensitive, but were also limited to those with low parameter correlations.

\begin{figure}
\includegraphics[width=\linewidth]{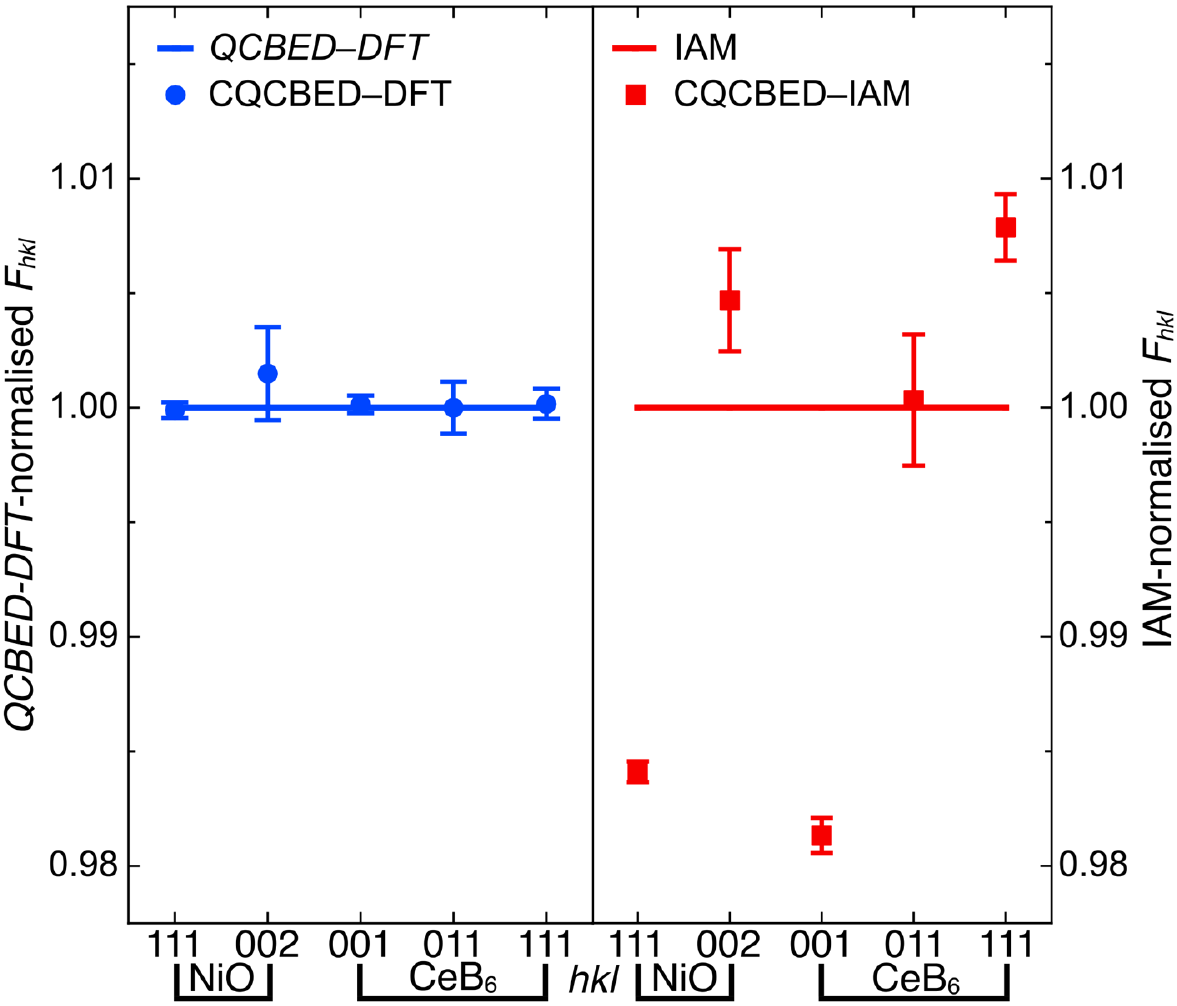}
\caption{(color online) A comparison of electron density structure factors, $F_{hkl}$, refined by CQCBED with different reference electronic structure models for NiO and CeB$_{6}$.  All points represent the refined $F_{hkl}$ normalized with respect to the corresponding model values.  The horizontal lines represent each model where \emph{QCBED-DFT} corresponds to DFT [APW + lo:  GGA(PBE) + $U$, ($U$ = 7.4 eV \& $J$ = 0.95 eV for NiO), ($U$ = 3 eV, $J$ = 0 eV \& $x$ =0.19925 for CeB$_{6}$)], as determined by our \emph{QCBED-DFT} refinements, and the IAM is the current \emph{International Union of Crystallography} standard \cite{DoyleTurnerIAM,IntTabCryst1974,IntTabCryst2006}.  All $F_{hkl}$ correspond to the temperatures of the experiments and were converted from $V_{hkl}$ by the Mott-Bethe formula \cite{Mott1930,Bethe1930}.  See Table S11 in \cite{SupplMats} for the actual values of $V_{hkl}$ and $F_{hkl}$.}
\label{comparisonfig}
\end{figure}

Figure \ref{comparisonfig} (see also Table S11 in \cite{SupplMats}) plots the CQCBED-DFT- and CQCBED-IAM-refined electron density structure factors, $F_{hkl}$, converted from potential structure factors, $V_{hkl}$, using the Mott-Bethe formula \cite{Mott1930,Bethe1930} and normalized by their modelled counterparts at the temperarures of the CBED experiments, $T$ = 110 K and $T$ = 298 K for NiO and CeB$_{6}$ respectively.  The horizontal lines at 1 on the ordinate axes represent the \emph{QCBED-DFT}-determined ${\rho}(\bf{r})$ (blue) and IAM ${\rho}(\bf{r})$ (red).  The large deviations of the IAM-normalized, CQCBED-IAM-refined $F_{hkl}$ from unity indicate the relative magnitudes of bonding effects on each of the structure factors.  In contrast, the \emph{QCBED-DFT}-normalized, CQCBED-DFT-refined $F_{hkl}$ for both NiO and CeB$_{6}$ have not departed from the \emph{QCBED-DFT}-modelled ${\rho}(\bf{r})$.  This is evidence that at least for NiO and CeB$_{6}$, GGA(PBE) + $U$ [$U_{NiO} = 7.4 \pm 0.6$ eV \& $J = 0.95$ eV; $U_{CeB_{6}} = 3.0 \pm 0.6$ eV \& $J = 0$ eV] generates electron densities that are as close to reality in these materials as may be determined from experimental CBED patterns.

The uncertainties in the CQCBED-DFT-refined structure factors are consistently smaller than those of CQCBED-IAM, with improvements in precision as much as three-fold in some cases (see also Table S11 in \cite{SupplMats}).  Replacing the IAM with a more accurate DFT model is bound to improve the accuracy of the electron scattering calculations within QCBED, yielding more consistent matches of calculated and experimental CBED patterns and therefore, reduced parameter uncertainties.

The demonstrated ability to ascertain the accuracy of modelled electron densities and the ability to refine DFT model parameters using \emph{QCBED-DFT} points to the possibility of developing parametrized density functionals guided by \emph{QCBED-DFT}.  Matching a 3-dimensional electron density is a more robust constraint than matching energies.  Therefore, \emph{QCBED-DFT} may be useful in searching for the exact density functional.

\bigskip
\bigskip

\begin{acknowledgments}
We dedicate this paper to the late Prof. A. W. S. Johnson who helped procure the CeB$_{6}$ single crystal (with Mr. P. Hanan, whom we also thank).  Prof. Johnson was very influential in this research.  We thank Dr. H. Cheng for help in preparing the NiO specimen.  We thank Prof. J. -M. Zuo for sharing his \emph{RefineCB5} code, which forms a significant component of the \emph{QCBED-DFT} package.  We are grateful to Dr. A. E. Smith for guidance and training in the use of the \emph{Wien2k} DFT package.  Many thanks to Prof. J. Etheridge FAA who reviewed our manuscript and gave us valuable advice.  The authors acknowledge the instruments and scientific and technical assistance at the Monash Centre for Electron Microscopy, Monash University, a node of Microscopy Australia, established under the Commonwealth Government's National Collaborative Research Infrastructure Strategy (NCRIS).  In particular, we thank A/Prof. L. Bourgeois and A/Prof. M. Weyland for microscopy training, expertise and advice.  We are very grateful to Dr. T. Liu for practical help with computational issues.  P. N. thanks the Australian Research Council for funding (FT110100427).
\end{acknowledgments}


\providecommand{\noopsort}[1]{}\providecommand{\singleletter}[1]{#1}%

\end{document}